\documentclass[12pt,prd]{revtex4}

\usepackage{amsmath}
\usepackage{amssymb}
\usepackage[dvips]{graphicx,color}

\begin{document}

\hfill\parbox{4cm}
{
 BNL-93821-2010-JA\\
 YITP-10-62\\
}

\title{
Classification of Minimally Doubled Fermions
}

\author{Michael Creutz}
\email{creutz@bnl.gov}
\affiliation{Physics Department, Brookhaven National Laboratory, 
Upton, NY 11973, USA}

\author{Tatsuhiro Misumi}
\email{misumi@yukawa.kyoto-u.ac.jp}
\affiliation{Yukawa Institute for Theoretical Physics, Kyoto University,
         Kyoto 606-8502, Japan}

\begin{abstract}
We propose a method to control the number of species of lattice
fermions which yields new classes of minimally doubled lattice
fermions.  We show it is possible to control the number of species by
handling $O(a)$ Wilson-term-like corrections in fermion actions, which
we will term ``Twisted-ordering Method".  Using this method we obtain
new minimally doubled actions with one exact chiral symmetry and exact
locality.  We classify the known minimally doubled fermions into two
types based on the locations of the propagator poles in the Brillouin
zone.
\end{abstract}

\maketitle

\newpage


\section{Introduction}
\label{sec:Intro}
Since the dawn of lattice field theory \cite{Wil}, the doubling problem
of fermions has been one of the notorious obstacles to QCD
simulations.  The famous no-go theorem of Nielsen and
Ninomiya \cite{NN} most clearly reveals the fate of lattice fermions.
It states that lattice fermion actions with chiral symmetry, locality
and other common features must produce degrees of freedom in multiples
of two in a continuum limit.  In contrast, phenomenologically there
exist only three quarks with masses below the underlying scale of
QCD.  By now several fermion constructions to bypass the no-go theorem
have been developed, although all of them have their individual
shortcomings.  For example, the explicit chiral symmetry breaking with
the Wilson fermion approach \cite{Wil} results in an additive mass
renormalization, which in turn requires a fine-tuning of the mass
parameter for QCD simulations.  Domain-wall fermion \cite{Kap, FuSh}
and overlap fermion \cite{GW, Neu} do not possess exact locality,
leading to rather expensive simulation algorithms.  These approaches
attempt to realize single fermionic degrees of freedom by breaking the
requisite conditions for the no-go theorem.

On the other hand, there is another direction to approach numerical
simulations.  According to \cite{KarS}, Hypercubic symmetry and
reflection positivity of actions result in $2^{d}$ species of fermions
where $d$ stands for the dimension.  Thus it is potentially possible
to reduce the number of species by breaking hypercubic symmetry
properly.  Actually, the staggered fermion approach \cite{KS, Suss,
  Sha}, with only 4 species of fermions does this and possesses
flavored-hypercubic symmetry instead.  However this requires rooting
procedures for the physical $2$ or $(2+1)$-flavor QCD simulation,
which is known to mutilate certain processes.

An interesting goal is a fermion action with only $2$ species, the
minimal number required by the no-go theorem.  More than 20 years ago,
Karsten and Wilczek proposed such a minimally doubled action \cite{Kar,
  Wilc}.  Recently one of the authors of the present paper
\cite{Creutz1, Bori, Creutz2} has also proposed a two-parameter class
of fermion actions, inspired by the relativistic condensed matter
system, graphene \cite{Cas}.  These minimally doubled fermions all
possess one exact chiral symmetry and exact locality.  As such they
should be faster for simulation, at least for two-flavor QCD, than
other chirally symmetric lattice fermions.

However it has been shown \cite{Bed1, Bed2, KM1, KM2} that we need to
fine-tune several parameters for a continuum limit with these actions.
This is because they lack sufficient discrete symmetry to prohibit
redundant operators from being generated through loop
corrections \cite{Cichy, Cap1, Cap2, Cap3}.  Thus the minimally doubled
fermions have not been extensively used so far.  Nevertheless, there
is the possibility to apply them to simulations if one can efficiently
perform the necessary fine-tuning of parameters.  To understand how
far we can reduce the number of fine-tuning parameters, it should be
useful to investigate and classify as many minimally doubled fermions
as possible.

In this paper we propose a systematic method to reduce the number of
doublers, which we term the ``twisted-ordering method".  
In this way we obtain new classes of minimally doubled fermions.  
We study symmetries of these fermion actions 
and show $CP$ invariance and $Z_{4}$ symmetry.  It is also pointed
out that they all also require fine-tuning of several parameters for a
correct continuum limit because of lack of sufficient discrete
symmetries.  We classify the known minimally doubled actions into two
types.  By this classification we can derive several further minimally
doubled actions deductively.

In Sec.~\ref{sec:TOM} we present the twisted-ordering method and
derive a new class of minimally doubled fermion actions.  In
Sec.~\ref{sec:MDF} we show that two classes of minimally doubled
actions arise from the original twisted-ordering actions.  In
Sec.~\ref{sec:Sym} we investigate the symmetries of these fermion
actions.  In Sec.~\ref{sec:CL} we discuss two classifications of 
minimally doubled fermions.  Section \ref{sec:summary} is devoted 
to a summary and discussion.


\section{Twisted-ordering Method}
\label{sec:TOM}
In this section we propose a systematic way of controlling the number
of species of lattice fermions within the requirement of
Nielsen-Ninomiya's no-go theorem.  We will firstly discuss the
$2$-dimensional case to give an intuitive understanding of this
mechanism.  Then we will go on to the $4$-dimensional case and show we
can construct a new minimally doubled action by this method.

\subsection{$2$-dimensional case}
\label{sec:2dTOM}

Let us begin with the following simple $2$-dimensional Dirac operator
in momentum space with $O(a)$ Wilson-like terms.
\begin{align}
D(p)\,\,=\,\,  &( \sin p_{1} \,+\, \cos p_{1}\,-\,1)\, i \gamma_{1}
\nonumber\\
+ &( \sin p_{2} \,+\, \cos p_{2} \,-\, 1 )\, i \gamma_{2},
\label{2dUT}
\end{align}
where $p_{1}$, $p_{2}$ and $\gamma_{1}$, $\gamma_{2}$ stand for
$2$-dimensional momenta and Gamma matrices respectively.  Here the
deviation from the usual Wilson action is that the $O(a)$ terms are
accompanied by Gamma matrices.  There are four zeros of this Dirac
operator at 
$(\tilde{p}_{1},\tilde{p}_{2})=(0,0),(0,\pi/2),(\pi/2,0),(\pi/2,\pi/2)$.
This means that the number of the species is the same as for naive
fermions.

\begin{figure}
\centering
\includegraphics[height=7cm]{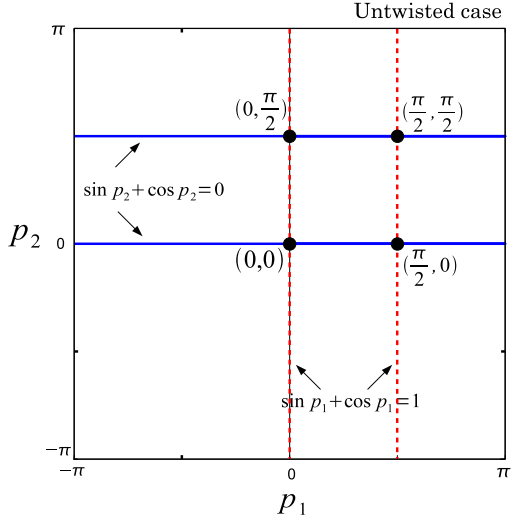} 
\caption{Zeros for the $2$-dimensional untwisted case are depicted
  within the Brillouin zone.  Red-dotted and blue-solid lines 
  stand for zeros of
  the coefficients of $\gamma_{1}$ and $\gamma_{2}$ respectively in
  Eq.~(\ref{2dUT}). Black points stand for the four zeros of the Dirac
  operator. }
 \label{UT}
\end{figure}

\begin{figure}
\centering
\includegraphics[height=7cm]{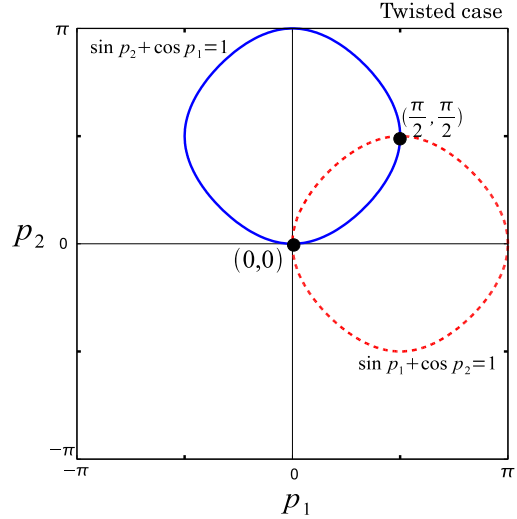} 
\caption{Zeros for the $2$-dimensional twisted case is depicted within
  the Brillouin zone.  Red-dotted and blue-solid curves 
  stand for zeros of the
  coefficients of $\gamma_{1}$ and $\gamma_{2}$ respectively in
  Eq.~(\ref{2d1T}). Black points stand for two zeros ($0,0$)
  ($\pi/2,\pi/2$) of the Dirac operator .}
 \label{1T}
\end{figure}

Next we ``twist'' the order of the $O(a)$ terms, or equivalently, permute
$\cos p_{1}$ and $\cos p_{2}$ to give
\begin{align}
D(p)\,\,=\,\,  &( \sin p_{1} \,+\, \cos p_{2}\,-\,1)\, i \gamma_{1}
\nonumber\\
+ &( \sin p_{2} \,+\, \cos p_{1} \,-\, 1 )\, i \gamma_{2}.
\label{2d1T}
\end{align}
Here only two of the zeros ($(0,0)$ and $(\pi/2,\pi/2)$) remain and
the other two are eliminated.  Thus the number of species becomes two,
the minimal number required by the no-go theorem.  As seen from this
example, twisting of the order of $O(a)$ terms reduces the number of
species.  We call this method ``twisted-ordering" in the rest of this
paper.  We depict the appearance of zeros in the Brillouin zone for
the above two cases in Fig.~\ref{UT} and \ref{1T}.
The red-dotted and blue-solid
curves represent the zeros of the coefficients of $\gamma_{1}$ and
$\gamma_{2}$ respectively.  These curves intersect at only two points
for the twisted case in Fig.~\ref{1T}.

Now consider excitations and the physical ($\gamma_{5}$)'s at the two zeros.  
Here the physical $\gamma_{5}$ at a zero is defined 
as the product of four coefficient matrices of momenta in the excitation 
as we will concretely show later.  
Now the Dirac operator is expanded about the zeros as
\begin{align}
D^{(1)}(q)\,\,&=\,\, i \gamma_{1}q_{1}  + \, i \gamma_{2}q_{2}\,+\, O(q^{2}),
\label{2d1T-Exp1}
\\
D^{(2)}(q)\,\,&=\,\, -i \gamma_{1}q_{2}  - \, i \gamma_{2}q_{1}\,+\, O(q^{2}).
\label{2d1T-Exp2}
\end{align}
Here we expand with respect to $q_{\mu}$ defined as
$p_{\mu}=\tilde{p}_{\mu}+q_{\mu}$ and we denote the two expansions as
$D^{(1)}$ for the zero ($0,0$) and $D^{(2)}$ for the other at
$(\pi/2,\pi/2)$.  Here momentum bases at the zero $(0,0)$ are given by
${\bf b}^{(1)}_{1}=(1,0)$ and ${\bf b}^{(1)}_{2}=(0,1)$ while those at
the other $(\pi/2,\pi/2)$ are given by ${\bf b}^{(2)}_{1}=(0,-1) $ and
${\bf b}^{(2)}_{2}=(-1,0)$.  This means excitations from the two zeros
describe physical fermions on the orthogonal lattice.  And the
$\gamma_{5}$ at $\tilde{p}=(0,0)$, which we denote as $\gamma_{5}^{(1)
}$, is given by
\begin{equation}
\gamma_{5}^{(1)}\,=\,\gamma_{1}\gamma_{2},
\end{equation} 
while the $\gamma_{5}$ at $\tilde{p}=(\pi/2,\pi/2)$, 
which we denote as $\gamma_{5}^{(2)}$, is given by
\begin{equation}
\gamma_{5}^{(2)}\,=\,\gamma_{2}\gamma_{1}\,=\,-\gamma_{5}^{(1)}.
\end{equation}
This sign change between the species is a typical relation between two
species with minimal doubling since Nielsen-Ninomiya's no-go theorem 
requires fermion pairs to possess chiral charges with opposite signs.
 
\subsection{$4$-dimensional case}
\label{sec:4dTOM}

Now let us go to the $4$-dimensional twisted-ordering method.  Here we
again start with the following simple action with $O(a)$ Wilson-like
terms
\begin{align}
D(p)\,\,=\,\,  &( \sin p_{1} \,+\, \cos p_{1}\,-\,1)\, i \gamma_{1}
\nonumber\\
+ &( \sin p_{2} \,+\, \cos p_{2} \,-\, 1 )\, i \gamma_{2}
\nonumber\\
+ &( \sin p_{3} \,+\, \cos p_{3} \,-\, 1 )\, i \gamma_{3}
\nonumber\\
+ &( \sin p_{4} \,+\, \cos p_{4} \,-\, 1 )\, i \gamma_{4}.
\label{4dUT}
\end{align}
Here each of the four momentum components can take either $0$ or
$\pi/2$ for zeros of this Dirac operator.  It means the associated
fermion action gives $16$ species doublers, the same as for naive
fermions.

Then we firstly consider the case with twisting the order of $O(a)$
terms once
\begin{align}
D(p)\,\,=\,\,  &( \sin p_{1} \,+\, \cos p_{1}\,-\,1)\, i \gamma_{1}
\nonumber\\
+ &( \sin p_{2} \,+\, \cos p_{2} \,-\, 1 )\, i \gamma_{2}
\nonumber\\
+ &( \sin p_{3} \,+\, \cos p_{4} \,-\, 1 )\, i \gamma_{3}
\nonumber\\
+ &( \sin p_{4} \,+\, \cos p_{3} \,-\, 1 )\, i \gamma_{4},
\label{4d1T}
\end{align}
where we permute the order of $\cos p_{3}$ and $\cos p_{4}$.
Here we find that there are $8$ zeros of the Dirac operator at
\begin{equation}
\tilde{p}_{1} = 0\,\,{\rm or}\,\,\pi/2,\,\,\,\,\,
\tilde{p}_{2} = 0\,\,{\rm or}\,\,\pi/2,\,\,\,\,\,
(\tilde{p}_{3},\tilde{p}_{4}) = (0,0)\,\,{\rm or}\,\,(\pi/2,\pi/2).
\label{Z-4d1T}
\end{equation}
The associated action yields $8$ species of doublers.  As seen from
this example, the twisted-ordering method reduces the number of
species also in $4$-dimensions.  In this case twisting is performed
only once, and the number of species is reduced by a factor of two.

Secondly we permute $\cos p_{1}$ and $\cos p_{2}$ in addition to $\cos
p_{3}$ and $\cos p_{4}$ to give
\begin{align}
D(p)\,\,=\,\,  &( \sin p_{1} \,+\, \cos p_{2}\,-\,1)\, i \gamma_{1}
\nonumber\\
+ &( \sin p_{2} \,+\, \cos p_{1} \,-\, 1 )\, i \gamma_{2}
\nonumber\\
+ &( \sin p_{3} \,+\, \cos p_{4} \,-\, 1 )\, i \gamma_{3}
\nonumber\\
+ &( \sin p_{4} \,+\, \cos p_{3} \,-\, 1 )\, i \gamma_{4}.
\label{4d2T}
\end{align}
In this case there are only $4$ zeros at
\begin{equation}
(\tilde{p}_{1},\tilde{p}_{2}) = (0,0)\,\,{\rm or}\,\,(\pi/2,\pi/2)\,\,\,\,\,
(\tilde{p}_{3},\tilde{p}_{4}) = (0,0)\,\,{\rm or}\,\,(\pi/2,\pi/2).
\label{Z-4d2T}
\end{equation}
Thus the associated action yields $4$ species of doublers.  This
example indicates that more twisting results in additional reduction
of species.  In these two examples, Eqs.~(\ref{4d1T}) and
(\ref{4d2T}), twisting is performed partially, and so we call this
kind of procedures ``partially-twisted-ordering".

Finally we consider a case that the order of $O(a)$ terms are
maximally twisted: We permute the $\cos$ terms in a cyclic way to give
\begin{align}
D(p)\,\,=\,\,  &( \sin p_{1} \,+\, \cos p_{2}\,-\,1)\, i \gamma_{1}
\nonumber\\
+ &( \sin p_{2} \,+\, \cos p_{3} \,-\, 1 )\, i \gamma_{2}
\nonumber\\
+ &( \sin p_{3} \,+\, \cos p_{4} \,-\, 1 )\, i \gamma_{3}
\nonumber\\
+ &( \sin p_{4} \,+\, \cos p_{1} \,-\, 1 )\, i \gamma_{4}.
\label{4dMT}
\end{align}
In this case there are only $2$ zeros at
\begin{equation}
(\tilde{p}_{1},\tilde{p}_{2},\tilde{p}_{3},\tilde{p}_{4})
=(0,0,0,0),\,\,\,(\pi/2,\pi/2,\pi/2,\pi/2).
\label{Z-4dMT}
\end{equation}
We have confirmed there are no other real zeros of this operator
numerically, and we will discuss an analytical proof in the next
section.  There are only two species in this case, the minimal number
required by the no-go theorem.  This is a new type of minimally
doubled fermion on the orthogonal lattice.

Now we study the excitations and $\gamma_{5}$'s at the two zeros in a
similar way to the $2$-dimensional case,
\begin{align}
D^{(1)}(q)\,\,&=\,\, i \gamma_{1}q_{1} + i \gamma_{2}q_{2} +
i\gamma_{3}q_{3 } + i\gamma_{4}q_{4} \,+\, O(q^{2}),
\label{4dMT-Exp1}\\
D^{(2)}(q)\,\,&=\,\, -i \gamma_{1}q_{2} - i \gamma_{2}q_{3} -
i\gamma_{3}q_{4 } - i\gamma_{4}q_{1} \,+\, O(q^{2}),
\label{4dMT-Exp2}
\end{align}
where we expand the Dirac operator (\ref{4dMT}) around zeros with
respect to $q_{\mu}$ defined as $p_{\mu}=\tilde{p}_{\mu}+q_{\mu}$ and
we denote the two expansions as $D^{(1)}$ for the zero ($0,0,0,0$) and
$D^{(2) }$ for the other $(\pi/2,\pi/2,\pi/2,\pi/2)$.  Momentum bases
at the zero $(0,0,0,0)$ are given by ${\bf b}^{(1)}_{1}=(1,0,0, 0)$,
${\bf b}^{(1)}_{2}=(0,1,0,0)$, ${\bf b}^{(1)}_{3}=(0,0,1,0)$ and ${\bf
  b}^{(1)}_{4}=(0,0,0,1)$ while those at the other zero at
$(\pi/2,\pi/2,\pi/2,\pi/2)$ are given by ${\bf b}^{(2
  )}_{1}=(0,0,0,-1)$, ${\bf b}^{(2)}_{2}=(-1,0,0,0)$, ${\bf
  b}^{(2)}_{3}=(0,-1,0,0)$ and ${\bf b}^{(2)}_{4}=(0,0,-1,0)$.  This
means excitations from the two zeros describe physical fermions on the
orthogonal lattice again.  And the $\gamma_{5}$ at
$\tilde{p}=(0,0,0,0)$, which we denote as $\gamma_{5}^ {(1)}$, is
given by
\begin{equation}
\gamma_{5}^{(1)}\,=\,\gamma_{1}\gamma_{2}\gamma_{3}\gamma_{4},
\end{equation} 
while the $\gamma_{5}$ at $\tilde{p}=(\pi/2,\pi/2,\pi/2,\pi/2)$, which
we denote as $\gamma_{5}^{(2)}$ is given by
\begin{equation}
\gamma_{5}^{(2)}\,=\,\gamma_{4}\gamma_{1}\gamma_{2}\gamma_{3}\,
=\,-\gamma_{5}^{(1)}.
\end{equation}
This is a typical relation between the two species with 
minimal-doubling.

Gauging these theories is straightforward.  One merely inserts the gauge
fields as link operators in the hopping terms for the action in
position space.  Specifically, for the
twisted-ordering minimally doubled
fermion in position space we have
\begin{equation}
S\,\,=\,\, {1\over{2}}\sum_{n,\mu}\Big[ \bar{\psi}_{n}\,
  \gamma_{\mu}\, (U_{n,\mu}\psi_{n+\mu} -
  U^{\dag}_{n-\mu,\mu}\psi_{n-\mu}) \,+\, i \bar{\psi}_{n}\,
  \gamma_{\mu-1}\, (U_{n,\mu}\psi_{n+\mu}\, +\,
  U^{\dag}_{n-\mu,\mu}\psi_{n-\mu} \,-\, 2\psi_{n})\Big],
\label{ACTION}
\end{equation} 
where we define 
\begin{equation}
\mu-1 \equiv 
\left\{\begin{matrix}
		 1,2,3 & (\mu=2,3,4) \\
		4 & (\mu=1).				   
\end{matrix}\right.
\label{mu+1}
\end{equation}
We note that in the twisted-ordering method, we control the number of
species by choosing the extent of breaking of discrete symmetries of
the action.  Twisting of $O(a)$ terms leads to breaking of 
the hypercubic symmetry.  On the other hand it is
well-known that hypercubic symmetry of the lattice fermion action
requires $16$ species as shown in \cite{KarS}.  Thus it is
reasonable that more breaking of discrete symmetries can lead to less
doubling.


\section{Minimally-Doubled Fermions}
\label{sec:MDF}

\begin{figure}
\centering
\includegraphics[height=7cm]{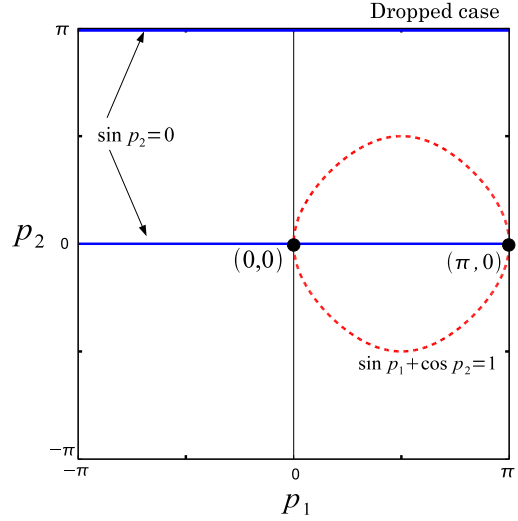} 
\caption{The two zeros for the Dirac operator (\ref{4dMT-D}) in the
  $2$-dimensional analogy.  Red-dotted and blue-solid lines stand for
  zeros of the coefficients of $\gamma _{1}$ and $\gamma_{2}$, namely
  ($\sin p_{1}+\cos p_{2}-1$) and $\sin p_{2}$ respectively.  Black
  dots show two zeros $(0,0)$ and $(\pi,0)$.}
 \label{T0}
\end{figure}

In this section we discuss two classes of minimally doubled actions
obtained from the original twisted-ordering action (\ref{4dMT}).  The
first one, which we call the ``dropped twisted-ordering action'', is
constructed as following: We drop one of $\cos p_{\mu}-1$ terms in the
Dirac operator (\ref{4dMT}), for example, drop $\cos p_{1}-1$ as
\begin{align}
D(p)\,\,=\,\,  &( \sin p_{1} \,+\, \cos p_{2}\,-\,1)\, i \gamma_{1}
\nonumber\\
+ &( \sin p_{2} \,+\, \cos p_{3} \,-\, 1 )\, i \gamma_{2}
\nonumber\\
+ &( \sin p_{3} \,+\, \cos p_{4} \,-\, 1 )\, i \gamma_{3}
\nonumber\\
+ &  \sin p_{4}\, i \gamma_{4}.
\label{4dMT-D}
\end{align}
This has two zeros given by
\begin{equation}
(\tilde{p}_{1},\tilde{p}_{2},\tilde{p}_{3},\tilde{p}_{4})\,
=\,(0,0,0,0),\,\,\,\,\, (\pi,0,0,0). 
\label{Z-4dMT-D}
\end{equation}
Thus the associated action is also a minimally doubled action.  Here
the two ($\gamma_{5}$)'s at the two zeros also have opposite signs
\begin{align}
\gamma_{5}^{(1)}\,&=\,\gamma_{1}\gamma_{2}\gamma_{3}\gamma_{4},
\nonumber\\
\gamma_{5}^{(2)}\,&=\,(-\gamma_{1})\gamma_{2}\gamma_{3}\gamma_{4}\,=\,-\gamma_{5}^{(1)},
\end{align}
where $\gamma_{5}^{(1)}$ corresponds to $(0,0,0,0)$ and $\gamma_{5}^{(2)}$ to $(\pi,0,0,0)$.
We can also confirm this action describes physical fermions on the orthogonal lattice
by looking into excitations from the zeros as in the previous section.
We depict the appearance of zeros for the two dimensional case of the
dropped twisted-ordering action in Fig.~\ref{T0}.  We see there are
only two zeros as $(0,0)$ and $(\pi,0)$.  The $4$-dimensional case is
similar.

Here we comment on definition of the equivalence of actions.  We can
convert one fermion action to an equivalent one with a different form
by momentum shift and rotation of Gamma matrices.  Here the momentum
shift means
\begin{equation}
p_{\mu} \,\,\,\to\,\,\, \pm p_{\mu}+k_{\mu},
\label{PSHIFT}
\end{equation}
where $k_{\mu}$ are some constant four vector.
The rotation of Gamma matrices means
\begin{equation}
\gamma_{\mu} \,\,\,\to\,\,\, C_{\mu\nu}\gamma_{\nu},
\label{GROT}
\end{equation}
where $C_{\mu\nu}$ is an orthogonal matrix.  Although by using such a
shift and rotation we can control the location of two zeros with the
distance and direction between them fixed, the original and
transformed actions are clearly equivalent.  Now we can easily see the
original (\ref{4dMT}) and dropped twisted-ordering (\ref{4dMT-D})
actions never translate into Borici \cite{Bori}
or Karsten-Wilczek \cite{Kar, Wilc} actions 
through these two procedures.  
It means our actions are independent and new ones.

Next we consider a second generalization
where we turn on a parameter in the action (\ref{4dMT}) as following,
\begin{align}
D(p)\,\,=\,\,  
  &( \sin p_{1} \,+\, \cos p_{2} \,-\, \alpha )\, i \gamma_{1}
\nonumber\\
+ &( \sin p_{2} \,+\, \cos p_{3} \,-\, \alpha )\, i \gamma_{2}
\nonumber\\
+ &( \sin p_{3} \,+\, \cos p_{4} \,-\, \alpha )\, i \gamma_{3}
\nonumber\\
+ &( \sin p_{4} \,+\, \cos p_{1} \,-\, \alpha )\, i \gamma_{4}.
\label{4dMT-a}
\end{align}
Here we replace unity with a positive parameter $\alpha$ in the
constant terms of the operator.  We will also call this one the
twisted-ordering fermion action in the rest of this paper.  The
$\alpha=1$ case corresponds to the original case.  Now, in order to
investigate the parameter range within which ``minimal-doubling" is
realized, we define $\sin p_{1}=x,\,\sin p_{2}=y,\,\sin p_{3}=z,\,\sin
p_{4}=w$ and rewrite the equations for zeros in terms of these
variables
\begin{align}
x \pm \sqrt{1-y^2} &= \alpha, \nonumber\\ 
y \pm \sqrt{1-z^2} &=
\alpha, \nonumber\\ 
z \pm \sqrt{1-w^2} &= \alpha, \nonumber\\ 
w \pm
\sqrt{1-x^2} &= \alpha.
\label{4dMT-x1}
\end{align}
To make it easier to see we rewrite this as
\begin{align}
x^2 + y^2 -2\alpha x  &= 1- \alpha^2, 
\nonumber\\
y^2 + z^2 -2\alpha y  &= 1- \alpha^2, 
\nonumber\\
z^2 + w^2 -2\alpha z  &= 1- \alpha^2, 
\nonumber\\
w^2 + x^2 -2\alpha w  &= 1- \alpha^2,
\label{4dMT-x2}
\end{align}
where in principle there are $16$ complex solutions.  The number of
solutions of this system is the same as the original ones
(\ref{4dMT-a})(\ref{4dMT-x1}).  Thus we can study the range of the
parameter for minimal-doubling by looking into how many real solutions
this system of equations has.  By eliminating $y$, $z$ and $w$, we
obtain the following equation for $x$
\begin{equation}
\alpha^{80}(2x^{2}-2\alpha x+\alpha^2-1)^{8}\,=\,0.
\label{4DEQ-Xonly}
\end{equation}
where we can confirm this by using software such as Maxima. 
This means, except for $\alpha=0$, there are only the two
solutions
\begin{equation}
x=y=z=w=\,{\alpha\pm\sqrt{2-\alpha^2}\over{2}}.
\end{equation}
From this $\alpha$ should be below $\alpha=\sqrt{2}$ so that the
solutions are real.  For $\alpha=0$ the above equation
(\ref{4DEQ-Xonly}) is satisfied by any value of $x$, which means that
zeros appear as a one-dimensional curve, not a point.  We will discuss
this in detail later.  We find there are two real solutions in the
range $0<\alpha<\sqrt{2}$, which corresponds to a pair of minimally
doubled species.  This is the range of the parameter $\alpha$ for
minimal-doubling.  We can also verify this minimal-doubling range
numerically.

What is significant for $\alpha=0$ is that zeros of the Dirac operator
(\ref{4dMT-a}) appear as an one-dimensional curve, not a point, in the
momentum space.  We can confirm this as following.  Equations for
zeros of the Dirac operator (\ref{4dMT-a}) at $\alpha=0$ are given by
\begin{align}
\sin p_1  &+  \cos p_2 = 0, \nonumber\\
\sin p_2  &+  \cos p_3 = 0, \nonumber\\
\sin p_3  &+  \cos p_4 = 0, \nonumber\\ 
\sin p_4  &+  \cos p_1 = 0.
\label{4DEQ}
\end{align}
These equations are satisfied by
\begin{align}
p_2 &= \pm (p_1 + \pi/2), \nonumber\\
p_3 &= \pm (p_2 + \pi/2), \nonumber\\
p_4 &= \pm (p_3 + \pi/2), \nonumber\\
p_1 &= \pm (p_4 + \pi/2),
\label{4DSOL}
\end{align}
where we ignore the arbitrariness of adding $\pm 2\pi$.
As seen from these relations, the solution appears
as an one-dimensional curve.
For example, you find the following curve satisfies the above equations:
\begin{equation}
(\tilde{p}_1, \tilde{p}_2, \tilde{p}_3, \tilde{p}_4)\,\,=\,\,
  (\tilde{p},\,\,-{\pi\over{2}}-\tilde{p},\,\,\tilde{p},\,
  \,-{\pi\over{2}}-\tilde{p}),
\label{4DZEROS}
\end{equation}
where $\tilde{p}$ can take an arbitrary value.  We can also understand
this from the fact that (\ref{4dMT-x2}) reduces to three independent
equations for $\alpha=0$.  This emergence of a line solution indicates
that the excitation around zeros for a small $\alpha$ becomes
incompatible with a Lorentz-covariant or a Dirac form.  Due to these
nonlinear corrections, we conclude that we should use this action only
with $\alpha$ substantially away from $\alpha\sim 0$.  This property
at $\alpha=0$ is also the case for any even number of dimensions.

\begin{figure}[tbp]
\centering
\includegraphics[height=7cm]{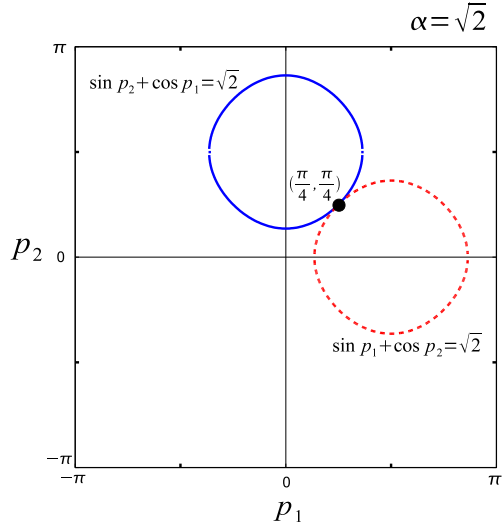} 
\caption{A $2$d analogy of two zeros at the upper limit $\alpha=\sqrt{2}$ in (\ref{4dMT-a}).
Red-dotted and blue-solid lines stand for zeros of the coefficients of $\gamma_{1}$ and $\gamma_{2}$,
namely ($\sin p_{1}+\cos p_{2}-\sqrt{2}$) and ($\sin p_{2}+\cos p_{1}-\sqrt{2}$) respectively.
Two zeros collide with each other and reduce to a double zero shown as a black point, 
whose excitation is not Lorentz-covariant.}
 \label{UPPER}
\end{figure}

\begin{figure}[tbp]
\centering
\includegraphics[height=7cm]{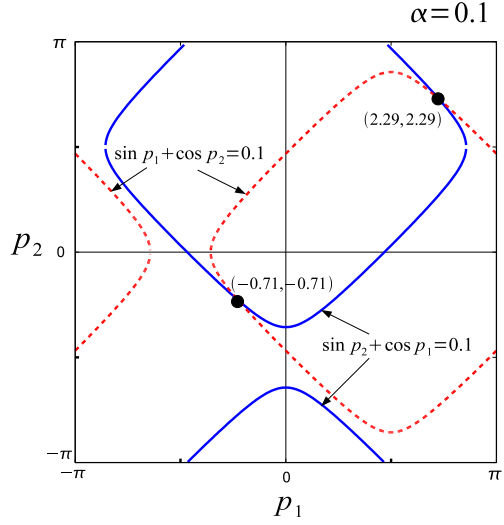} 
\caption{A $2$d analogy of two zeros for the lower limit $\alpha=0.1$ in (\ref{4dMT-a}).
Red-dotted and blue-solid lines stand for zeros of the coefficients of $\gamma_{1}$ and $\gamma_{2}$,
namely ($\sin p_{1}+\cos p_{2}-0.1$) and ($\sin p_{2}+\cos p_{1}-0.1$) respectively.
Two zeros shown as black points go further to each other with $\alpha$ decreasing.
There are nonlinear corrections to the dispersion relation, 
which leads to Lorentz-non-covariant excitations.}
 \label{LOWER1}
\end{figure}

\begin{figure}[tbp]
\centering
\includegraphics[height=7cm]{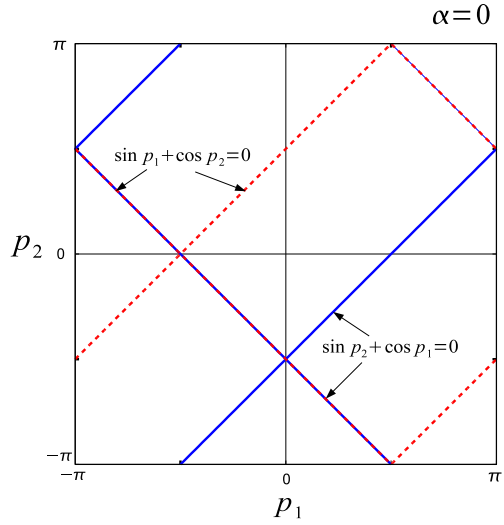} 
\caption{A $2$d analogy of two zeros for the lower limit $\alpha=0$ in (\ref{4dMT-a}).
Red-dotted and blue-solid lines stand for zeros of the coefficients of $\gamma_{1}$ and $\gamma_{2}$,
namely ($\sin p_{1}+\cos p_{2}$) and ($\sin p_{2}+\cos p_{1}$) respectively.
There are piled lines composed of red and blue lines, which stand for one-dimensional line solutions.}
\label{AL0}
\end{figure}

We can visualize the allowed range of the parameter $\alpha$ from the
$2$-dimensional analogy.  In the $2$d case the upper limit of
$\sqrt{2}$ is illustrated in Fig.~\ref{UPPER}.  For $\alpha=\sqrt{2}$
the two zeros join to a double zero, where the excitation is
not Lorentz-covariant.  For a lower limit, the $2$d analogy shows
``Minimal-doubling" persists as long as $\alpha$ is positive.  We
depict $2$d cases for $\alpha=0.1$ and $\alpha=0$ in Fig.~\ref{LOWER1}
and \ref{AL0}.  For $\alpha=0$ the solution appears as an
one-dimensional line, thus there is no simple pole in the propagator.

We conclude that the parameter range for minimal-doubling in both $2$
and $4$ dimensions is given by
\begin{equation}  
0\,<\,\alpha \,<\, \sqrt{2}.
\label{MDR}
\end{equation}
In this range two zeros of the $4$d Dirac operator, 
which we denote as $\tilde{p}_{\mu}^{(1)}$
and $\tilde{p}_{\mu}^{(2)}$, are given by
\begin{align}
\tilde{p}_{\mu}^{(1)}\,&=\,
\arcsin \left({\alpha-\sqrt{2-\alpha^2}\over{2}} \right) 
\,=\, \arcsin A,  
\label{Z-4dMT-a1}
\\
\tilde{p}_{\mu}^{(2)}\,&=\,
\arcsin \left({\alpha+\sqrt{2-\alpha^2}\over{2}} \right) 
\,=\, \arcsin \sqrt{1-A^{2}},
\label{Z-4dMT-a2}
\end{align}
where for convenience we define a parameter $A$
as $A\equiv (\alpha-\sqrt{2-\alpha^2})/2$.
These two zeros reduce to Eq.~(\ref{Z-4dMT}) with $\alpha=1$.
Excitations from these zeros are given by
\begin{align}
D^{(1)}(q)\,=\, &i\sqrt{1-A^{2}}(\gamma_{1}q_{1} +\gamma_{2}q_{2} 
+ \gamma_{3}q_{3} + \gamma_{4}q_{4})
-iA(\gamma_{1}q_{2}+\gamma_{2}q_{3}
+ \gamma_{3}q_{4}+\gamma_{4}q_{1}) \,+\, O(q^{2}),
\label{4dMT-a-Exp1}\\
D^{(2)}(q)\,=\, &iA(\gamma_{1}q_{1} +\gamma_{2}q_{2} 
+ \gamma_{3}q_{3} + \gamma_{4}q_{4})
-i\sqrt{1-A^{2}}(\gamma_{1}q_{2}+\gamma_{2}q_{3}
+ \gamma_{3}q_{4}+\gamma_{4}q_{1}) \,+\, O(q^{2}),
\label{4dMT-a-Exp2}
\end{align}
where $O(q^{2})$ corrections are not negligible 
for a small $\alpha$ as we have discussed. 
Momentum bases at the zero $p_{\mu}^{(1)}$ are given by 
${\bf b}^{(1)}_{1}=(\sqrt{1-A^{2}},0,0,-A)$,
${\bf b}^{(1)}_{2}=(-A,\sqrt{1-A^{2}},0,0)$, 
${\bf b}^{(1)}_{3}=(0,-A,\sqrt{1-A^{2}},0)$ 
and ${\bf b}^{(1)}_{4}=(0,0,-A,\sqrt{1-A^{2}})$ 
while those at the other zero are given by 
${\bf b}^{(2)}_{1}=(A,0,0,-\sqrt{1-A^{2}})$, 
${\bf b}^{(2)}_{2}=(-\sqrt{1-A^{2}},A,0,0)$, 
${\bf b}^{(2)}_{3}=(0,-\sqrt{1-A^{2}},A,0)$ 
and ${\bf b}^{(2)}_{4}=(0,0,-\sqrt{1-A^{2}},A)$.  
They are clearly non-orthogonal for cases of $\alpha\not=1$
thus the associated lattices are non-orthogonal as with the
Creutz fermion \cite{Creutz1}.

We note we have confirmed the parameter range for minimal-doubling
(\ref{MDR}) is also the case with other even dimensions
($d=2,4,6,8,10$).  From this fact we can anticipate this is a
universal property for this class of actions for any even dimensions.

Here we comment on the ranges of the parameter for minimal-doubling in
odd dimensions.  For example, in the $3$d case, we can rewrite
analogous equations for zeros of the operator with $\sin p_1 =x$,
$\sin p_2 = y$ and $\sin p_3 =z$ as
\begin{align}
x^2 + y^2 -2\alpha x  &= 1- \alpha^2, 
\nonumber\\
y^2 + z^2 -2\alpha y  &= 1- \alpha^2, 
\nonumber\\
z^2 + x^2 -2\alpha z  &= 1- \alpha^2.
\nonumber\\ 
\label{3DEQ-X}
\end{align}
By eliminating $y$ and $z$,
we obtain the following equation for $x$,
\begin{align}
(2x^{2}&-2\alpha x+\alpha^2-1) (8 x^{6}-24\alpha x^{5}
+(36\alpha^2 -12)x^{4}-(36\alpha^{3}-24\alpha )x^{3}
\nonumber\\
&+(2\alpha^{4}-40\alpha^{2}+6)x^{2}
+(10\alpha^{5}+28\alpha^{3}-6\alpha )x +25\alpha^{6}+29\alpha^{4}
+11\alpha^{2}+1)\,=\,0.
\label{3DEQ-Xonly}
\end{align} 
As seen from this there are two real solutions similar with the $4$d
case for $\alpha<\sqrt{2}$ as
\begin{equation}
x=y=z=\,{\alpha\pm\sqrt{2-\alpha^2}\over{2}}.
\end{equation}
However, in this case we have six more complex solutions which come
from the other factor (\ref{3DEQ-Xonly}).  The question is how many
real solutions this equation has.  
We have studied it numerically and found six real solutions exist 
below a critical value of $\alpha$ as $\alpha=0.28...$
while there are six complex solutions above it.
Thus the minimal-doubling parameter range
in the $3$d case is given by
\begin{equation}
0.28... < \alpha < \sqrt{2}.
\end{equation}
In this case the lower limit is non-zero while the upper limit is the
same as even dimensional cases.  Indeed we can numerically confirm
this kind of nonzero lower limit is also the case with any odd
dimensional twisted-ordering action.  We can explain the reason that
there is such difference between the even and odd dimensional cases as
following: As seen from (\ref{3DEQ-X}), there are $d$ independent
equations even for $\alpha=0$ in the ($d=2n+1$)-dim cases
($n=1,2,3,...$) while there are only $d-1$ equations in the even
dimensions for $\alpha=0$.  This means that the Dirac operator can
have $2^{d}$ zeros for $\alpha=0$ in the odd dimensional cases while
zeros appear as a curve in even dimensions.  Actually, as seen from
(\ref{3DEQ-Xonly}), there are $2^{d}$ independent real solutions for
$\alpha=0$ in the odd dimensions.  Since only two zeros exist for
$\alpha=1$ also for odd dimensions, there must be a critical point
between $\alpha=0$ and $\alpha=1$ in odd dimensions.


\section{Symmetries}
\label{sec:Sym}

In the previous section we have studied two classes of
minimally doubled actions in (\ref{4dMT-D}) and (\ref{4dMT-a}), both
of which are obtained from the twisted-ordering action in (\ref{4dMT}).
In this section we will discuss discrete symmetries of these actions
and redundant operators generated by loop corrections.

Firstly we note both of the actions possess some common properties
with other minimally doubled fermions: ``gamma-five Hermiticity", 
``discrete translation invariance", 
``flavor-singlet $U(1)_{V}$" and ``flavor-nonsinglet
$U(1)_{A}$." The latter is the exact chiral symmetry preventing
additive mass renormalization for the neutral pion.  
In addition to them, they have exact
locality and gauge invariance when gauged by link variables.  We
expect every sensible minimally doubled fermion to possess these basic
properties.

On the other hand, Discrete symmetries associated with permutation of
the axes and $C$, $P$ or $T$ invariance depends on a class of the
actions.  In other words, we can identify minimally doubled actions by
these discrete symmetries.  Here we show discrete symmetries which the
dropped twisted-ordering action possesses:
\vspace{3mm}

1. $CP$, 
\vspace{2mm}

2. $T$, 
\vspace{2mm}

3. $Z_{2}$ associated with two zeros.
\vspace{2mm}

Here we take the $p_{4}$ direction as time.  We also write the
symmetries which the full twisted-ordering action possesses:
\vspace{3mm}

1. $CPT$, 
\vspace{2mm}

2. $Z_{4}$ associated with axes permutation,
\vspace{2mm}

3. $Z_{2}$ associated with two zeros.
\vspace{2mm}

There is also the possibility to utilize the above $Z_{2}$ symmetry in
order to recover flavored-$C$, $P$ or $T$ symmetry.  Actually
Borici-Creutz action has a flavored-$C$ symmetry \cite{Bed1}.  
However we do not further discuss here details of this point, 
to which a future work will be devoted.

The dropped twisted-ordering action has $CP$ and $T$ invariance.
We consider what redundant operators can be generated radiatively in
this action following the prescription shown in \cite{Bed1}.  
The relevant and marginal operators generated through
loop corrections are given by
\begin{align}
\mathcal{O}_{3}^{(1)} &= \bar{\psi}i\gamma_{1}\psi, \nonumber\\
\mathcal{O}_{3}^{(2)} &= \bar{\psi}i\gamma_{2}\psi, \nonumber\\
\mathcal{O}_{3}^{(3)} &= \bar{\psi}i\gamma_{3}\psi, 
\label{RELE}
\end{align}
and
\begin{align}
\mathcal{O}_{4}^{(1)} &= \bar{\psi}\gamma_{\mu}D_{\mu}\psi, \nonumber\\
\mathcal{O}_{4}^{(2)} &= \bar{\psi}\gamma_{1}D_{1}\psi, \nonumber\\
\mathcal{O}_{4}^{(3)} &= \bar{\psi}\gamma_{2}D_{2}\psi, \nonumber\\
\mathcal{O}_{4}^{(4)} &= \bar{\psi}\gamma_{3}D_{3}\psi, \nonumber\\
\mathcal{O}_{4}^{(5)} &= F_{\mu\nu}F_{\mu\nu}, \nonumber\\
\mathcal{O}_{4}^{(6)} &= F_{1\nu}F_{1\nu}, \nonumber\\
\mathcal{O}_{4}^{(7)} &= F_{2\nu}F_{2\nu}, \nonumber\\
\mathcal{O}_{4}^{(8)} &= F_{3\nu}F_{3\nu}.
\label{MAR}
\end{align}
These operators are allowed by the symmetries for possessed by the
action. Note that the marginal operators include a renormalization 
of the speed of light, both for the fermions and the gluons.

On the other hand, the case of twisted-ordering action is similar to
that of Borici-Creutz action which has higher symmetry $S_{4}$ than
$Z_{4}$ \cite{Bed1}.  It is unlikely that there are more redundant
operators generated in the twisted-ordering action than those of
Borici-Creutz one shown in \cite{Bed1}.  However in the future we need
to confirm that this reduction of symmetry does not affect the number
of redundant operators.  Then we will know whether or not this class
is more useful than the Borici-Creutz class.


\section{Classification}
\label{sec:CL}
In this section we discuss a classification of the known minimally
doubled fermions.  So far we have seen four variations on minimally
doubled fermions: Karsten-Wilczek, Borici-Creutz, twisted-ordering and
dropped twisted-ordering fermions.  Here we classify these actions
into two types: One of them, which includes Karsten-Wilczek and
dropped twisted-ordering actions, is given by
\begin{equation} 
D(p)\,=\, \sum_{\mu}i\gamma_{\mu}\sin p_{\mu}\,
+\, \sum_{i,j} i\gamma_{i}R_{ij}(\cos p_{j}-1),
\label{GENE1}
\end{equation}
where we take $i=1,2,3$ as $i$ while $j=2,3,4$ as $j$. 
The point is that indices $i$ and $j$
are staggered.  
The different
actions in this class depend on the choice for the matrix $R$.
For example, consider the following $R$'s
\begin{align}
R&=\left( 
\begin{matrix}
1\,\, & 1\,\, & 1 \\
0\,\, & 0\,\, & 0 \\
0\,\, & 0\,\, & 0 \\
\end{matrix}
\right),
\label{RKW}
\\
R&=\left( 
\begin{matrix}
1\,\, & 0\,\, & 0 \\
0\,\, & 1\,\, & 0 \\
0\,\, & 0\,\, & 1 \\
\end{matrix}
\right),
\label{RT}
\\
R&=\left( 
\begin{matrix}
1\,\, & 1\,\, & 0 \\
0\,\, & 0\,\, & 1 \\
0\,\, & 0\,\, & 0 \\
\end{matrix}
\right).
\label{RNEW}
\end{align}
In the case of (\ref{RKW}), the general form (\ref{GENE1}) reduces to
Karsten-Wilczek action as following,
\begin{align}
D(p)\,\,=\,\,  
&( \sin p_{1}+\cos p_{2}+\cos p_{3}+\cos p_{4}-3)\, i \gamma_{1}
\nonumber\\
+ &\sin p_{2}\, i \gamma_{2}
\nonumber\\
+ &\sin p_{3}\, i \gamma_{3}
\nonumber\\
+ &\sin p_{4}\, i \gamma_{4}.
\label{KW}
\end{align}
Here you can also add an overall factor as a parameter to $R$.
Actually the original Karsten-Wilczek action includes such a parameter
$\lambda$ in front of $O(a)$ terms with a minimal-doubling range of
the parameter $\lambda >1/2$.

In the case of (\ref{RT}), this reduces to the dropped
twisted-ordering action
\begin{align}
D(p)\,\,=\,\,  &( \sin p_{1} \,+\, \cos p_{2}\,-\,1)\, i \gamma_{1}
\nonumber\\
+ &( \sin p_{2} \,+\, \cos p_{3} \,-\, 1 )\, i \gamma_{2}
\nonumber\\
+ &( \sin p_{3} \,+\, \cos p_{4} \,-\, 1 )\, i \gamma_{3}
\nonumber\\
+ &  \sin p_{4}\, i \gamma_{4},
\end{align}
where you can also add an overall parameter to $R$.

Finally, the action for the case (\ref{RNEW}) is a new possibility,
given by
\begin{align}
D(p)\,\,=\,\,  
&(  \sin p_{1}+\cos p_{2}+\cos p_{3}-2)\, i \gamma_{1}
\nonumber\\
+ &(\sin p_{2}+\cos p_{4}-1)\, i \gamma_{2}
\nonumber\\
+ &\sin p_{3}\, i \gamma_{3}
\nonumber\\
+ &\sin p_{4}\, i \gamma_{4}.
\label{NEW}
\end{align}

We see that there are many options associated with possible $R$'s.
One common property with fermions obtained in this way is that two
zeros are separated along a single lattice axis and given by
$(0,0,0,0)$ and $(\pi,0,0,0)$.  One significance about the general
form (\ref{GENE1}) is that a coefficient of at least one Gamma matrix
has no $(\cos p_{\mu}-1)$ term.  It is also notable that, in a
coefficient of each gamma matrix, a momentum component $p_{\mu}$
associated with a sine term differs from the component in any cosine
term.  These two points seem to be essential to minimal-doubling.
Here we note minimal-doubling still persists if we add a $(\cos
p_{1}-1)$ term to any actions obtained from (\ref{GENE1}).  We can
change the location of the two zeros by adding such a term with a
parameter.

Here let us comment on a sufficient condition of $R$ 
for minimal-doubling.
We found minimal-doubling is realized 
when you take the following class of $R$'s:
$R$'s composed by three column vectors as $R=(v_{1}, v_{2}, v_{3})$ 
with each vector $v_{i}$ ($i=1,2,3$) taken 
as $(1,0,0)^{T}$, $(0,1,0)^{T}$ or $(0,0,1)^{T}$.
All the examples we have shown here satisfy this condition.
However we have not yet revealed a necessary condition for this.
Our future work will be devoted to revealing such a condition.

The second class of actions includes the twisted-ordering fermion with
the $\alpha$ parameter and the Borici-Creutz actions.  A generalized
Dirac operator for this type is given by
\begin{equation}
D(p)\,\,=\,\, i\sum_{\mu}[ \gamma_{\mu}\sin (p_{\mu}+\beta_{\mu}) \,
-\, \gamma_{\mu}'\sin (p_{\mu}-\beta_{\mu}) ]\,-\, i\Gamma ,
\label{GENE2}
\end{equation} 
where $\gamma_{\mu}'=A_{\mu\nu}\gamma_{\nu}$ is another set of gamma
matrices where we define $A$ as an orthogonal matrix with some
conditions: At least one eigenvalue of $A$ should be $1$ and all four
components of the associated eigenvector should have non-zero values.
Here $\beta_{\mu}$ and $\Gamma$ has a relation with this $A$ as
$\Gamma=\sum_{\mu}\gamma_{\mu}\sin
2\beta_{\mu}=\sum_{\mu}\gamma_{\mu}'\sin 2\beta_{\mu}$, which means
$\sin 2\beta_{\mu}$ is an eigenvector of $A$ as $A_{\mu\nu}\sin
2\beta_{\nu}=\sin 2\beta_{\mu}$.  Thus once $A$ is fixed,
$\beta_{\mu}$ and $\Gamma$ are determined up to a overall factor of
$\sin 2\beta_{\mu}$.  By imposing these conditions on $A$,
$\beta_{\mu}$ and $\Gamma$, the action (\ref{GENE2}) can be a
minimally doubled action.  In such a case $\beta_{\mu}$ indicates
locations of two zeros as $\tilde{p}_{\mu}=\pm\beta_{\mu}$.  Adjusting
$\beta$, we can control the locations of the zeros.

Note that the first and second terms in (\ref{GENE2}) are nothing but
naive fermion actions.  We can eliminate some species by combining two
naive actions with different zeros in one action.

Now let us show this action includes minimally doubled actions.
Firstly this general form in (\ref{GENE2}) reduces to Borici action by
choosing $A$, which is given by
\begin{align} 
\gamma'_{\mu} \,&=\, A_{\mu\nu} \gamma_{\nu},
\nonumber\\
A \,&=\,  {1\over{2}}\left( 
\begin{matrix}
-1 & 1 & 1 & 1 \\
1 & -1 & 1 & 1 \\
1 & 1 & -1 & 1 \\
1 & 1 & 1 & -1
\end{matrix}
\right),
\label{GENE2-B}
\end{align}
where we fix $\beta_{\mu}=\pi/4$ thus $\Gamma$ is given by
$\Gamma=\sum_{\mu}\sin(\pi/2)\gamma_{\mu}=\sum_{\mu}\gamma_{\mu}$.

This general form (\ref{GENE2}) also yields the twisted-ordering
action.  For this we take the following matrix $A$
\begin{align} 
\gamma'_{\mu} \,&=\, A_{\mu\nu} \gamma_{\nu},
\nonumber\\
A \,&=\,  \left( 
\begin{matrix}
0 & 0 & 0 & 1 \\
1 & 0 & 0 & 0 \\
0 & 1 & 0 & 0 \\
0 & 0 & 1 & 0
\end{matrix}
\right), 
\label{GENE2-T}
\end{align}
fix $\beta_{\mu}=\pi/4$, and $\Gamma$ is given by
$\Gamma=\sum_{\mu}\gamma_{\mu}$.  Then the general form reduces to the
twisted-ordering action with $\alpha=1$.  We can also obtain
the $\alpha\not=1$ case by choosing other values of $\beta_{\mu}$.  For
$0<\alpha<\sqrt{2}$ we take $0<\beta_{\mu}<\pi/2$ along with which an
overall factor for $\Gamma$ changes.

We note this general form (\ref{GENE2}) also includes non-minimally
doubled actions like the partially twisted-ordering actions in
(\ref{4d1T}) and (\ref{4d2T}).  Here let us consider
\begin{align} 
\gamma'_{\mu} \,&=\, A_{\mu\nu} \gamma_{\nu},
\nonumber\\
A \,&=\,  \left( 
\begin{matrix}
0 & 1 & 0 & 0 \\
1 & 0 & 0 & 0 \\
0 & 0 & 0 & 1 \\
0 & 0 & 1 & 0
\end{matrix}
\right),
\label{GENE2-PT}
\end{align}
and fix $\sin 2\beta_{\mu}$ where $\Gamma$ is given by
$\Gamma=\sum_{\mu}\gamma_{\mu}$.  This clearly gives the
partially-twisted-ordering action in (\ref{4d2T}).

Finally we have classified the known minimally doubled fermions into
the two types.  Now we can derive a lot of varieties from these
general forms deductively as we have shown in (\ref{NEW}).  However
the second type of general actions (\ref{GENE2}) is not restricted
only to minimally doubled fermions.  We may be able to restrict it by
imposing other conditions.  We note the two fermion actions which we
derived from the twisted-ordering method, namely twisted-ordering and
dropped twisted-ordering, belong to different types of minimally
doubled fermions although they look similar to each other.


\section{Summary}
\label{sec:summary}
In this paper we study a variety of new classes of minimally doubled
fermions and a classification of all the known cases.  In
Sec.~\ref{sec:TOM} we propose a systematic method to reduce the number
of species, the ``twisted-ordering method".  In this method we can choose
$2$, $4$, $8$ and $16$ as the number of species by controlling
discrete symmetry breaking.  By using this method we obtain new classes
of minimally doubled fermions, which we call twisted-ordering and
dropped twisted-ordering actions.  In Sec.~\ref{sec:MDF} we study
these two classes in terms of the parameter range for
minimal-doubling.  In Sec.~\ref{sec:Sym} we study discrete symmetries
of these fermion actions and show twisted-ordering action has $Z_{4}$
symmetry while dropped twisted-ordering action possesses $CP$
invariance.  We also show they require fine-tuning of several
parameters for a correct continuum limit because of lack of sufficient
discrete symmetries in the known classes.  In Sec.~\ref{sec:CL} we
classify all the known minimally doubled actions into two types.  We
can derive several unknown minimally doubled actions by this
classification.

With several varieties of minimally doubled fermion actions available,
a goal is to apply these actions to numerical simulations and study
their relative advantages.  In particular it is important to
investigate how many fine-tuning parameters are required for a good
continuum limit and how difficult this tuning is.  This kind of study
was started in \cite{Bed1}.  Recently renormalization of the operators
has been also studied in \cite{Cap1, Cap2, Cap3}.  Both of these
studies focused on Karsten-Wilczek and Borici-Creutz fermions.  One
can now explore other varieties such as ones in this paper.  With a
better understanding of their properties, one can go on to larger
numerical calculations.


\begin{acknowledgments}
TM is supported by Grand-in-Aid 
for the Japan Society for Promotion of Science (JSPS) Research Fellows(No.\ 21-1226).
TM thanks Taro Kimura for stimulating discussions.  
MC is grateful to the Alexander von Humboldt Foundation for support for visits to
the University of Mainz.  This manuscript has been authored under
contract number DE-AC02-98CH10886 with the U.S.~Department of Energy.
Accordingly, the U.S. Government retains a non-exclusive,
royalty-free license to publish or reproduce the published form of
this contribution, or allow others to do so, for U.S.~Government
purposes.
\end{acknowledgments}



\begin{thebibliography}{99}
\bibitem{Wil}
K.~G.~Wilson, Phys.Rev.D {\bf 10},  2445 (1974). 

\bibitem{NN}
H.~B.~Nielsen and M.~Ninomiya, Nucl. Phys. B {\bf 185}, 20 (1981); 
Nucl. Phys. B {\bf 193} 173 (1981);
Phys. Lett. B {\bf 105} 219 (1981).

\bibitem{Kap}
D.~B.~Kaplan, Phys. Lett. B {\bf 288}, 342 (1992).

\bibitem{FuSh}
V.~Furuman and Y.~Shamir, Nucl. Phys. B {\bf 439}, 54 (1995).

\bibitem{GW}
P.~H.~Ginsparg and K.~G.~Wilson, Phys. Rev. D {\bf 25}, 2649 (1982).

\bibitem{Neu}
N.~Neuberger, Phys. Lett. B {\bf 427}, 353 (1998).  

\bibitem{KarS}
L.~H.~Karsten and J.~Smit, Nucl. Phys. B {\bf 183}, 103 (1981). 

\bibitem{KS}
J.~B.~Kogut and L.~Susskind, Phys. Rev. D {\bf 11}, 395 (1975).

\bibitem{Suss}
L.~Susskind, Phys. Rev. D {\bf 16}, 3031 (1977).

\bibitem{Sha}
H.~S.~Sharatchandra, H.~J.~Thun and P.~Weisz, Nucl. Phys. B {\bf 192}, 205 (1981). 

\bibitem{Kar}
L.~H.~Karsten, Phys. Lett. B {\bf 104}, 315 (1981). 

\bibitem{Wilc}
F.~Wilczek, Phys. Rev. Lett. {\bf 59}, 2397 (1987). 

\bibitem{Creutz1}
M.~Creutz, JHEP 0804, 017 (2008) [arXiv:0712.1201 [hep-lat]]. 

\bibitem{Bori}
A.~Borici, Phys. Rev. D {\bf 78}, 074504 (2008) [arXiv:0712.4401 [hep-lat]]; PoS LATTICE2008, (2008) [arXiv:0812.0092].

\bibitem{Creutz2}
M.~Creutz, PoS LATTICE2008, (2008) [arXiv:0808.0014]. 

\bibitem{Cas}
A.~H.~Castro Neto, F.~Guinea, N.~M.~R.~Peres, K.~S.~Novoselov,
A.~K.~Geim, Rev. Mod. Phys. 81, 109 (2009),
arXiv:0709.1163 [cond-mat.other]. 

\bibitem{Bed1}
P.~F.~Bedaque, M.~I.~Buchoff, B.~C.~Tiburzi and A.~Walker-Loud, Phys. Lett. B {\bf 662}, 449 (2008) [arXiv:0801.3361 [hep-lat]];
M.~I.~Buchoff, PoS LATTICE2008 (2008) [arXiv:0809.3943].

\bibitem{Bed2}
P.~F.~Bedaque, M.~I.~Buchoff, B.~C.~Tiburzi and A.~Walker-Loud, Phys. Rev. D {\bf 78}, 017502 (2008) [arXiv:0804.1145 [hep-lat]]. 

\bibitem{KM1}
T.~Kimura and T.~Misumi, to appear in Prog. Theor. Phys. [arXiv:0907.1371 [hep-lat]].

\bibitem{KM2}
T.~Kimura and T.~Misumi, Prog. Theor. Phys. {\bf 123}, 63 (2010) [arXiv:0907.3774 [hep-lat]].

\bibitem{Cichy}
K. Cichy, J. Gonzalez Lopez, K. Jansen, A. Kujawa and A. Shindler, Nucl. Phys. B {\bf 800}, 94 (2008) [arXiv:0802.3637 [hep-lat]]. 

\bibitem{Cap1}
S.~Capitani, J.~Weber, H.~Wittig, Phys. Lett. B {\bf 681}, 105 (2009) [arXiv:0907.2825].

\bibitem{Cap2}
S.~Capitani, J.~Weber, H.~Wittig, (2009) [arXiv:0910.2597].

\bibitem{Cap3}
S.~Capitani, M.~Creutz, J.~Weber, H.~Wittig, (2010) [arXiv:1006.2009].


\end{thebibliography}
\end{document}